# Imaging Hydrogen Bond in Real Space

Xiu CHEN[1,§], Shuyi LIU[1,§], Lacheng LIU[1], Xiaoqing LIU[1], Yingxing Cai[1], Nianhua LIU[1,2], Li WANG[1,2,*]

[1] *Department of Physics, Nanchang University, Nanchang 330031, P.R. China*

[2] *Nanoscience and Nanotechnology Laboratory, Institute for Advanced Study, Nanchang University, Nanchang 330031, P. R. China*

**Abstract:** Hydrogen bond is often assumed to be a purely electrostatic interaction between a electron-deficient hydrogen atom and a region of high electron density. Here, for the first time, we directly image hydrogen bond in real space by room-temperature scanning tunneling microscopy (STM) with the assistance of resonant tunneling effect in double barrier mode. STM observations demonstrate that the C=O:HO hydrogen bonds lifted several angstrom meters above metal surfaces appear shuttle-like features with a significant contrast along the direction connected the oxygen and hydrogen atoms of a single hydrogen bond. The off-center location of the summit and the variance of the appearance height for the hydrogen bond with scanning bias reveal that there are certain hybridizations between the electron orbitals of the involved oxygen and hydrogen atoms in the C=O:HO hydrogen bond.

*) liwang@ncu.edu.cn
§) Equal contribution

**One sentence summary:** Hydrogen bond has been imaged in real space for the first time and the formation of such hydrogen bond is attributed to hybridization of atomic orbitals rather than purely electrostatic interaction.

**Main text:** Hydrogen bonds (H-bonds) are momentous and ubiquitous interactions in chemistry and biology[1-2]. They are responsible for the structure and properties of water, an essential compound for life [3]. Moreover, H-bonds also play a key role in determining the three-dimensional shapes, chemical and physical properties, and functions of biomolecules, such as proteins and nucleic acid [4]. Hydrogen bond is proposed to be a purely and directionally electrostatic interaction between electron-deficient hydrogen and a region of high electron density by Pauling [5]. Recently, compton scattering in ordinary ice[6], nuclear magnetic resonance (NMR) measurements in nuclear acid base pairs[7], blue shift of the X-H stretching frequency on hydrogen bond formation[8], have given unambiguous evidences for incompleteness of electrostatic picture for hydrogen bond and raised arguments for the origin of hydrogen bonds [9, 10].

Although all the methods of physical chemistry, spectroscopy and diffraction can be used to recognize and study hydrogen bonds, the existence of hydrogen bonds is only inferred a posteriori from the spatial proximity and relative orientation of the hydrogen bond donor, the hydrogen, and the hydrogen bond acceptor once the structure of a molecule is solved by either X-ray crystallography or NMR. The information on H-bonds is derived indirectly from their profound influences and often leaves ambiguities. It is, therefore, desirable to get direct information on the properties of hydrogen bonds. Very few experimental techniques can provide direct information on hydrogen bonds, especially the spatial distribution and bonding nature of hydrogen bonds. The hydrogen bonds (O··H-C) within a self-assembly of 3,4,9,10-perlenetetracrboxylic dianhydride (PTCDA) were recently recorded as the thin lines between the areas with different contrasts by scanning tunneling hydrogen microscopy (STHM) [11]. Unfortunately, no any information on hydrogen bonds can be derived from the observation due to its unknown physical origin of the intermolecular bonding contrast for this technique.

Since scanning tunneling microscopy (STM) was invented [12], STM has been proved as a powerful tool with incredible resolution to image atoms, molecule and bond in real space[ 13-15]. In a STM image, the discriminability on different parts directly relies on the differences in the tunneling current between a tip and a surface during the tip scans across the surface under constant current mode. For a H-bonded self-assembly of molecules on a surface, the contribution of the H-bonds to the measured tunneling current are completely buried by the contribution from the underneath surface due to the weak nature of H-bonds, which makes it impossible to pinpoint the H-bonds directly by STM under conventional constant current or constant high modes. In order to overcome this disadvantage and make H-bonds show significant contrast to a supporting surface, we propose a new mode to visualize H-bonds without any modifications on STM hardware based on resonant tunneling effect through double barriers. In this so called "double barrier mode", an H-bond is lifted several angstrom above a metal surface in geometry and decoupled from the supporting surface in electronic structure, as shown in Fig.1a.

During the scanning process, there are two types of tunneling junctions between a metal tip and the metal surface: for the bare part of the surface, the tunneling junction can be ascribed as a single vacuum barrier between the tip and the surface; for the H-bond region, the junction is described as a vacuum barrier between the tip and the H-bond (B1), a H-bond well and a vacuum barrier between the H-bond and the metal surface (B2), as shown in Fig.1b. In order to simply the case and emphasize the role of the double-barrier, the transmission coefficient in such double-barrier configuration under one-dimensional model has been calculation as a function of the width of B2 ($d_{B2}$) and the energy of the injected electron. The situation where the $d_{B2}$ is zero represents that a hydrogen bond directly lies on a metal surface. As shown in Fig.1c, the transmission coefficient is significantly enhanced with the increase of $d_{B2}$ due to the presence of double barriers. The maximum transmission coefficient in double-barrier configuration with $d_{B2} = 0.2$ nm and E= 1.58 eV is close to 1, almost four orders magnitude larger than that for hydrogen bond lying on surface configuration.

In our study, C=O:HO hydrogen bond with an energy of 7.4 Kcal/mol [16] is chosen as an representative for moderate or normal hydrogen bonds which are believed to be mostly electrostatic[1,10]. In order to lift hydrogen bonds away from solid surface, 1,3-Adamantanedicarboxylic Acid (AA) molecules, one of Adamantane derivatives, are adopted as adsorbates. As shown in Fig.2a, the core of AA molecule, Adamantane, consists of ten carbon atoms in sp$^3$ form and two carboxylic acid groups are connected to two of three carbon atoms which are bonded to three neighbored carbon atoms. Once the AA molecules are deposited onto a Au(111) surface at room temperature, STM observations clearly demonstrates that these adsorbed molecules spontaneously form well ordered self-assembled structures on the surface, as shown in Fig.2b. The AA molecules appear as the bright triangle-shape or inverted triangle-shape features in the STM images and are arranged themselves into two kind of molecular rows labeled as A and B, respectively. The self-assembled structures of the AA molecules on Au(111) surface are consisted of the cross-arranged molecular chains A and B with the basic unit vectors of a=1.35 nm, b=0.56 nm. Fig.2c gives the proposed model for this self-assembled structure calculated in the frame of density function theory (DFT). The structural model clearly shows that the carboxylic acid groups of the AA molecules in a row form the hydrogen bonds with the carboxylic acid groups of the molecules in the neighbored rows. Such formation of hydrogen bonds between two AA molecules reduces the energy of 0.6 eV for one molecule. The distances between two molecules in the same molecular row and between two hydrogen-bonded molecules in the difference rows are 0.56 nm and 0.70 nm, which are in good agreement with the values found in our STM observations. In order to meet the requirements for the formation of hydrogen bonds between the AA molecules within the neighbored rows, the direction of the molecular arrangement for the AA molecules in a row must be opposite to that for the molecule in the neighbored row. Such molecular arrangements are also consistence with our STM observations that the AA molecules in a row are triangle-shape and the molecules in the neighbored rows are inverted triangle-shape.

It is worth noting that the distribution of the hydrogen bonds on this self-assembled monolayer appears zigzag under such molecular arrangement. From the side view of the structural model, the hydrogen bonds formed by the carboxylic acid groups are lifted to be ~0.40 nm away from the gold surface, which fits the experimental conditions to image the hydrogen bonds in the double barrier mode we proposed. As a fact, close examinations to Fig.2b clearly reveal that there is the presence of the dark brown line-like features denoted by white lines within the trenches between the molecular rows. In Fig.2d, the height distribution of the obtained STM image is rescaled to emphasize these line-like features and the corresponding molecular arrangement is also included to show the directions where the C=O:HO hydrogen bonds formed by the carboxylic acid groups are. It is unambiguous that the line-like features denoted by white dash lines well match the predicted directions of the C=O:HO hydrogen bonds. Moreover, the distribution of such line-features is indeed zigzag as what expected from the structural model in Fig.2c. Therefore, these line-like features within the trenches of the self-assembled structures are believed to represent the C=O:HO hydrogen bonds between the AA molecules, or in the other words, the hydrogen bonds between the AA molecules are indeed imaged as the line-like features with significant contrasts to the background in real space under the double barrier mode we proposed.

Since the C=O:HO hydrogen bonds between the AA molecules is ~0.2 nm below the top of the AA molecule, it is hard to quantitatively measure the properties of these hydrogen bonds due to the limited movement of the STM tip. Placing C=O:HO bonds on the top of an assembled structure is crucial for performing quantative measurements of the hydrogen bonds. Therefore, 1,1'-Ferrocenedicarboxylic Acid (FA), one of the sandwich compounds consisted of two cyclopentadienyl rings bound on opposite sides of a central Fe atom in Fig.3a, is chosen as adsorbate. When these organometallic molecules adsorb on a solid surface with their cyclopentadienyl rings parallel to the surface, the H-bonds formed between the COOH groups within the top cyclopentadienyl rings of two neighbored FA molecules can be lifted 0.36 nm above the solid surface and almost in the same plane as the top cyclopentadienyl rings of the molecules, which gives the possibility to directly measure the profile of hydrogen bonds. Fig.3b gives the typical STM image for FA molecules adsorbed on a Cu(100) surface at room temperature. The adsorbed FA molecules are spontaneously assembled into molecular chains along the [101] or [110] directions of the copper surface. High resolution STM image in Fig.3c show that these ordered molecular chains are actually described as grape-like structures. From our STM observations, it is clear that each of the bright round features within such grape-like structures actually represents a FA molecule adsorbed on the copper surface with its cyclopentadienyl rings parallel to the surface. The separation between two neighbored molecules in the same side and the opposite side are 0.73 nm and 1.03 nm, respectively. Fig.3d gives the proposed model for such grape-like assembled structures. DFT calculations reveal that the driving forces to form such ordered structures are the hydrogen bonds between the neighbored FA molecules and the formation of a hydrogen bond cause a reduce of an energy of 0.6 eV for one FA molecule. In Fig.3c, the DFT model for such assembled structure, a COOH group of a FA molecule at one side of grape-like feature forms C=O:HO or H:O-C hydrogen bonds with the COOH groups of two neighbored FA molecules at the opposite side of the grape-like feature. For example, the molecule labeled as A1 form C=O:HO hydrogen bonds denoted by yellow lines with the molecules labeled as B1 and B2 and at the same time the B2 molecule forms two hydrogen bonds with the A1 and A2 molecules. Therefore, the distribution of the C=O:HO bonds within the grape-like assembled structures is zigzag-like along the [101] or [110] surface directions. Close examinations to the high-resolution STM images in Fig.3d clearly reveal that the grape-like assembled structures do have zigzag-like "stems" to connect each of the FA molecules as DFT calculations predicted. Therefore, each line-like features denoted by yellow lines that form the observed "stem" actually represents single C=O:HO hydrogen bond, which further confirms that hydrogen bond can be imaged under the double barrier mode. Moreover, the brightness of such line-like features is almost the same as that for the FA molecule, suggesting that these hydrogen bonds are almost in the same plane as the top cyclopentadienyl ring of the FA molecule.

Further analysis on the C=O:HO bonds in the assembled FA chains reveal that the observed hydrogen bonds are not simple lines but actually they have symmetric shuttle-like distribution in space along the lines connected the negative centers (the oxygen atoms) and the positive centers (the hydrogen atoms), as shown in Fig.4a. Such spatial distribution of the observed hydrogen bonds are completely different from what found in assembled PDCTA structures on Au(111) surface observed by STHM: straight and thin lines without any spatial distribution on the boundary between two domains in different contrasts [11]. Whatever it is in the frame of Paul's theory or hybridization theory, hydrogen bonds must have certain spatial distributions as the electrostatic fields or hybridized electron orbitals involved in hydrogen bonds are spatial-related. Moreover, hydrogen bonds must be rotationally symmetric along their central axis due to the symmetry of their atomic arrangement. The images of the hydrogen bonds presented here are in good agreement with these predicted properties of hydrogen bonds, further confirming that the shuttle-like feature for hydrogen bond recorded in "double barrier mode"indeed reflects the real situation of a single C=O:HO hydrogen bond.

From the hypsographic map of the area highlighted in the white frame, the inset of Fig.4a, it is immediately realized that there is a summit within the area for a hydrogen bond and this summit does not locate in the line connected the oxygen atom and the hydrogen atom involved in a single hydrogen bond. For a single C=O:HO hydrogen bond, an coordinate to describe the location of the summit of this hydrogen bond is constructed in this way: the center of the line connected the hydrogen atom and oxygen atom is chosen as the origin point; the direction from the hydrogen atom to the oxygen atom and the direction perpendicular to the H-O line are adopted as the X and Y coordinate axes; as shown in the inset of Fig.4a. The statistic results over hundreds of C=O:HO hydrogen bonds at various areas of FA assembled structures in Fig. 4b reveal that the most likely position for the summit of C=O:HO hydrogen bond obviously deviates from the origin point, the center of the line between the hydrogen atom and the oxygen atom, and can be described with the coordinations (-0.42, -0.31). Figure 4c shows the arrangement for the hydrogen (cyan cirle) and the oxygen atom (red circle) of the OH group, the oxygen atom (red circle) of the C=O part and the summit of the hydrogen bond (cross feature) in the same coordinate system described the above. It is surprisedly found that the summit of the hydrogen bond actually locates in the line (dash line) connected the oxygen atom and the hydrogen atom in the OH group rather than the line between the hydrogen atom and the oxygen atom in the hydrogen bond, which evidently derives from what purely electrostatic interaction theory predicts and indicates the possibility for the occurrence of orbital hybridization within hydrogen bond.

In a STM image, the appearance height for an object is determined by both the topological height and the local density state of the electronic structure for such object. In other words, for a same object, its appearance height in the STM images could vary with the applied STM-sample bias, moreover, this difference of the appearance height only reflects the diversity of the local density state for various electronic orbitals since the geometrical structure of the concerned object remains unchangeable. Figure 4d shows the STM images recorded at various bias voltages for the same area of a FA assembled chain.

Although these STM images appear almost the same to another, the profile measurements along the directions across two opposite cyclopentadienyl rings and the middle H bond shown in Fig.4e demonstrate that there are dramatic changes on the appearance heights for the cyclopentadienyl rings of the molecules and the H bond. Under a scanning bias of -3V, the appearance height for the cyclopentadienyl ring is about 1.2Å and larger than that (~1.05Å) for the H bond; once the scanning bias is switched to be -0.25V, the height of the cyclopentadienyl ring (~1.1Å) is smaller than that (~1.3Å) for the H bond. The dependence of the appearance height of the cyclopentadienyl ring on the scanning bias in Fig.4e clearly show that the height of the ring first gradually decreases from ~1.2Å to ~0.9Å with the decrease of the bias from -3V to – 1.5V and then sostenuto increases to ~1.3Å with the further decrease of the bias. Such changes on the appearance height of the cyclopentadienyl ring with the bias reflect the differences on the local density states at various energy levels of the FA molecule rather than the actual changes on the geometry of the FA molecule. Interestingly, the dependence of the appearance height for H bond area on the scanning bias follows the same tendency as the height of the cyclopentadienyl ring does. It is reasonable to deduce that the changes on the height of the H bond are indeed caused by the differences on the local density state of the electronic orbitals within the H bond region. The presence of the local density state of electronic oritals within the H bond region is an evident sign for the occurrence of the hybridization between the oxygen atom and the hydrogen atom that forms this H bond. Therefore, the C=O:OH hydrogen bond observed is not originated from purely electrostatic interaction and there are certain hybridizations between the electron orbitals of the involved oxygen and hydrogen atoms.

We have directly imaged C=O:HO hydrogen bonds lifted from metal surface by STM in a double barrier mode. The spatial distribution of the C=O:HO hydrogen bonds in the high-resolution images and the dependence of the appearance height of hydrogen bonds with scanning bias reveal that the formation of such hydrogen bond is attributed to hybridizations between the electron orbitals of the involved oxygen and hydrogen atoms rather than purely electrostatic interaction. Visualization of weak bonds in double-barrier mode provides a feasible way to study the physical and chemical properties of weak bonds and throws a light on understanding the origin of such weak bonds.

**Ackonlegement:** This work was supported by National Key Basic Research Program of China (2013CB934200) and Natioanl Science Foundation of China (Nos. 10979015 and 50976048). L. Wang acknowledges the financial supports from the Program for New Century Execellent Talents in University, Ministry of Education of the People's Republic of China and Jiangxi Provincial "Ganpo Talentes 555 Projects".

**References:**

1. G. A. Jeffrey, *An Introduction to Hydrogen Bonding*, Oxford University Press, Oxford, 1997.

2. G. A. Jeffrey, W. Saenger, *Hydorgen Bonding in Biological Stuctures*, Springer, Berlin, 1991

3. S. Scheiner, *Hydrogen Bonding*, Oxford University Press, New York, 1997

4. P. Schanda, M. Huber, R. Verel, M. Ernst and B. H. Meier, Angew. Chem. Int. Ed. 48, 9322-9325 (2009)

5. L. Pauling, *The Nature of the Chemical Bond*, Cornell University Press, Ithaca, New York , 1960

6. E. D. Isaacs, A. Shukla, P. M. Platzman, D. R. Hamann, B. Barbiellini and C. A. Tulk, Physical Review Letters 82, 600-603 (1999)

7. A. J. Dingley and S. Grzesiek, J. Am. Chem. Soc. 120, 8293 (1998)

8. P. Hobza and A. Havla, Chem. Rev. 100, 4253 (2000)

9. G. R. Desiraju, Angew. Chem. Int. Ed. 50, 52-59 (2011); Acc. Chem. Res. 35, 565-573 (**2002)**

10. T. Steiner, Angew. Chem. Int. Ed. 41, 48-76 (2002)

11. C. Weiss, C. Wagner, R. Temirov and F. S. Tautz, J. Am. Chem. Soc. 132, 11864-11865 (2010)

12. G. Binning, H. Rohrer, Ch. Geber, and E. Weibel, *Physical Review Letters*, **49**, 57 (1982)

13. B. C. Schardt, S.-L. Yau, and F. Rinaldi, *Science*, **243**, 1050 (1989)

14. A. Zhao, Q. Li, L. Chen, H. Xiang, W. Wang, S. Pan, B. Wang, X. Xiao, J. Yang, J. G. Hou, Q. Zhu, Science 309, 1542-1544 (2005)

15. J. Repp, G. Meyer, S. Paavilainen, F. E. Olsson and M. Persson, Science 312, 1196-1199 (2006)

16. T. Neuheuser, B. A. Hess, C. Reutel, E. Weber, J. Phys Chem. 98, 6459 -6467 (1994)

**Figure captions:**

Figure 1 : Double-barrier mode for STM observations. (a) sketch for lifting hydrogen bond from solid surface; (b) diagrams for hydrogen bond lying on a solid surface and lifted away from a solid surface and the corresponding tunnelling barrier configurations; (c) the transmission coefficient thought turnneling barrier in one-dimensional model as a function of the width of the barrier between the hydrogen bond and the surface and the involved electron energy.

Figure 2: H-bonded self-assembly of 1,3-Adamantanedicarboxylic Acid (AA) on Au(111) surface. (a) molecular structure of AA molecules under topview and sideview configurations; (b) a typical STM image for self-assembled AA molecules, 5 nm x 5 nm, $V_t$=3 V, $I_t$=0.01 nA ; (c) structure model for the H-bonded self-assembled structure. The dark, white and red balls in the model represent carbon, hydrogen and oxygen atoms, respectively. The yellow balls are for gold atoms on the surface; (d) high-resolution STM image to show the H-bonds between two AA molecules, 2.5 nm x 2.5 nm, $V_t$=3 V, $I_t$=0.01 nA. The corresponding arrangement of AA molecule is also included for comparison.

Figure 3: H-bonded molecular chains of 1,1'-Ferrocenedicarboxylic Acid (FA) on Cu(100) surface. (a) molecular structure of FA molecules under topview and sideview configurations; (b) a typical STM image for AA molecular chains, 30 nm x 30 nm, $V_t$=3 V, $I_t$=0.01 nA ; (c) high-resolution STM image for a AA molecular chain, 3.2 nm x 3.2

nm, $V_t$=3 V, $I_t$=0.01 nA;(d) structure model for the H-bonded molecular chain. The dark, white and red balls in the model represent carbon, hydrogen and oxygen atoms, respectively. The green balls are for copper atoms on the surface.

Figure 4: Spatial distribution of individual C=O:HO bond within a FA molecular chain. (a) high-resolution STM image and corresponding molecular arrangement for FA molecular chain. 1.6 nm x 1.6 nm, $V_t$=3 V, $I_t$=0.01 nA; the inset shows the hypsographic map of the highlighted area; (b) the statistical distribution of the locations for the summits of the hydrogen bonds; (c) the arrangement for the positions of the oxygen atom and the hydrogen atom in the OH group and the oxygen atom in the C=O group in a C=O:HO hydrogen bond. (d) STM images taken at -2850mV, -850mV and -125mV; (e) the profile measurements of the appearance height at scanning biases of -2850mV, -850mV and -125mV; (f) the appearance height for FA molecule and the hydrogen bond at various bias.

Figure 1

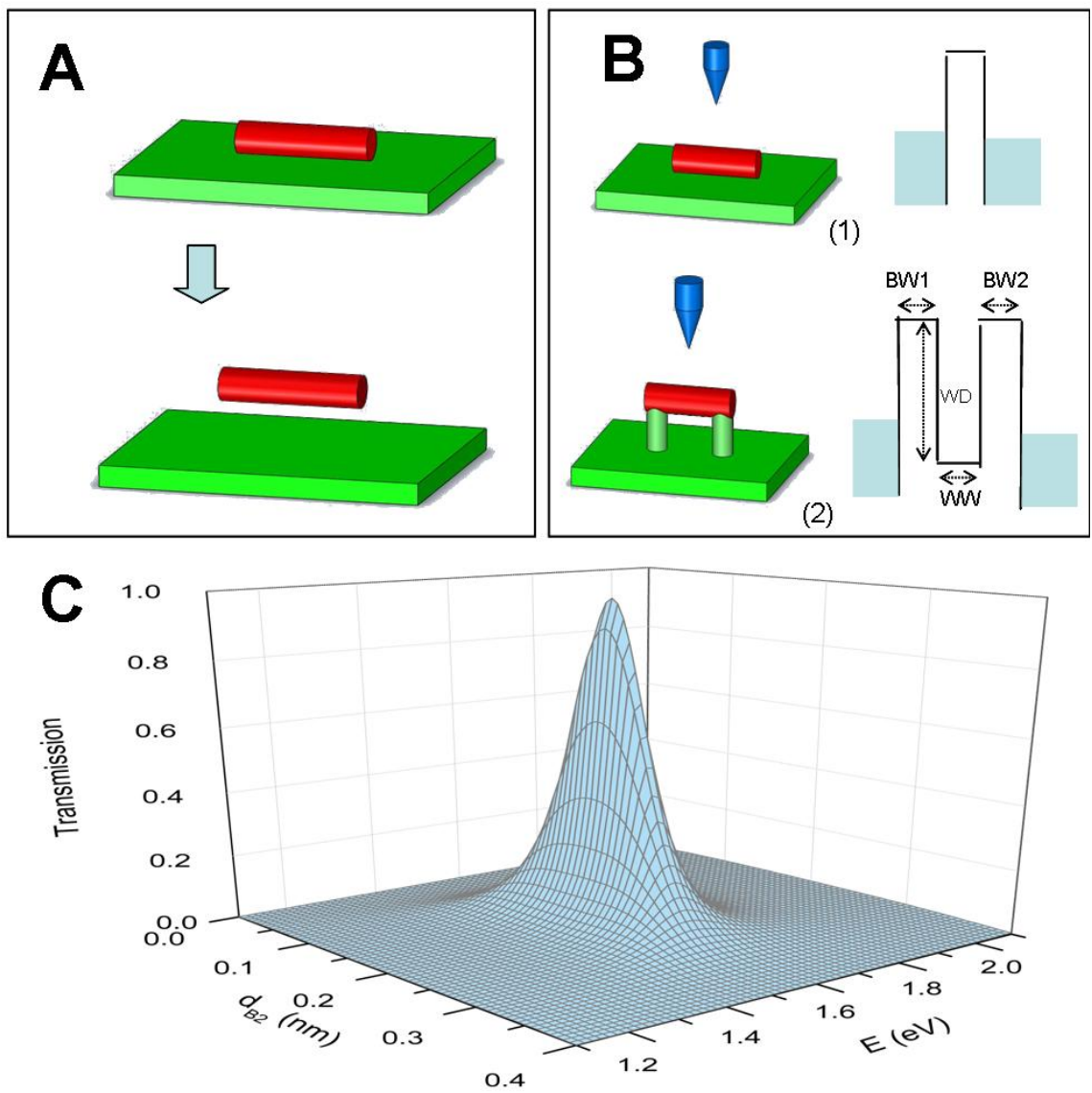

Figure 2

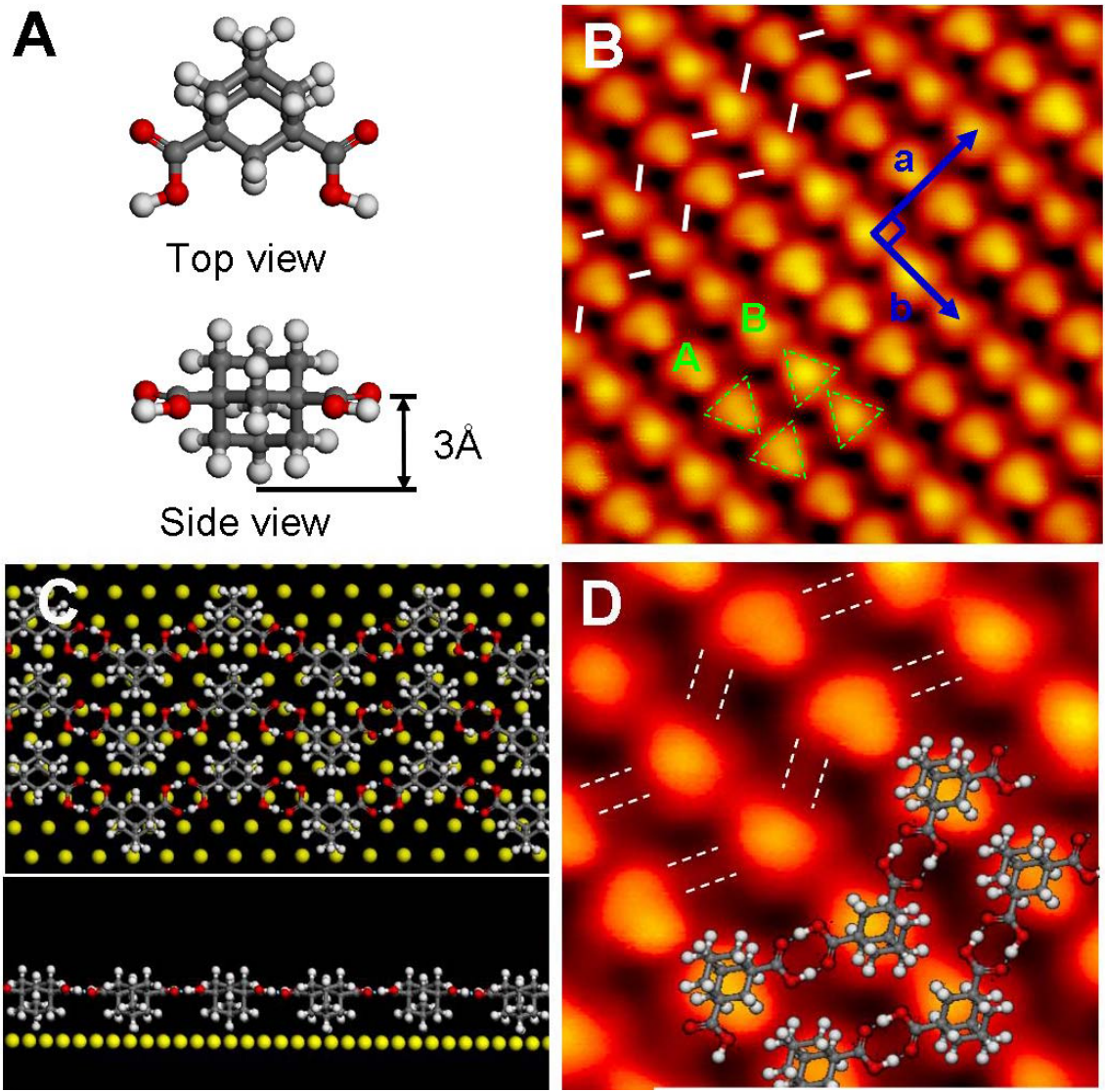

Figure 3

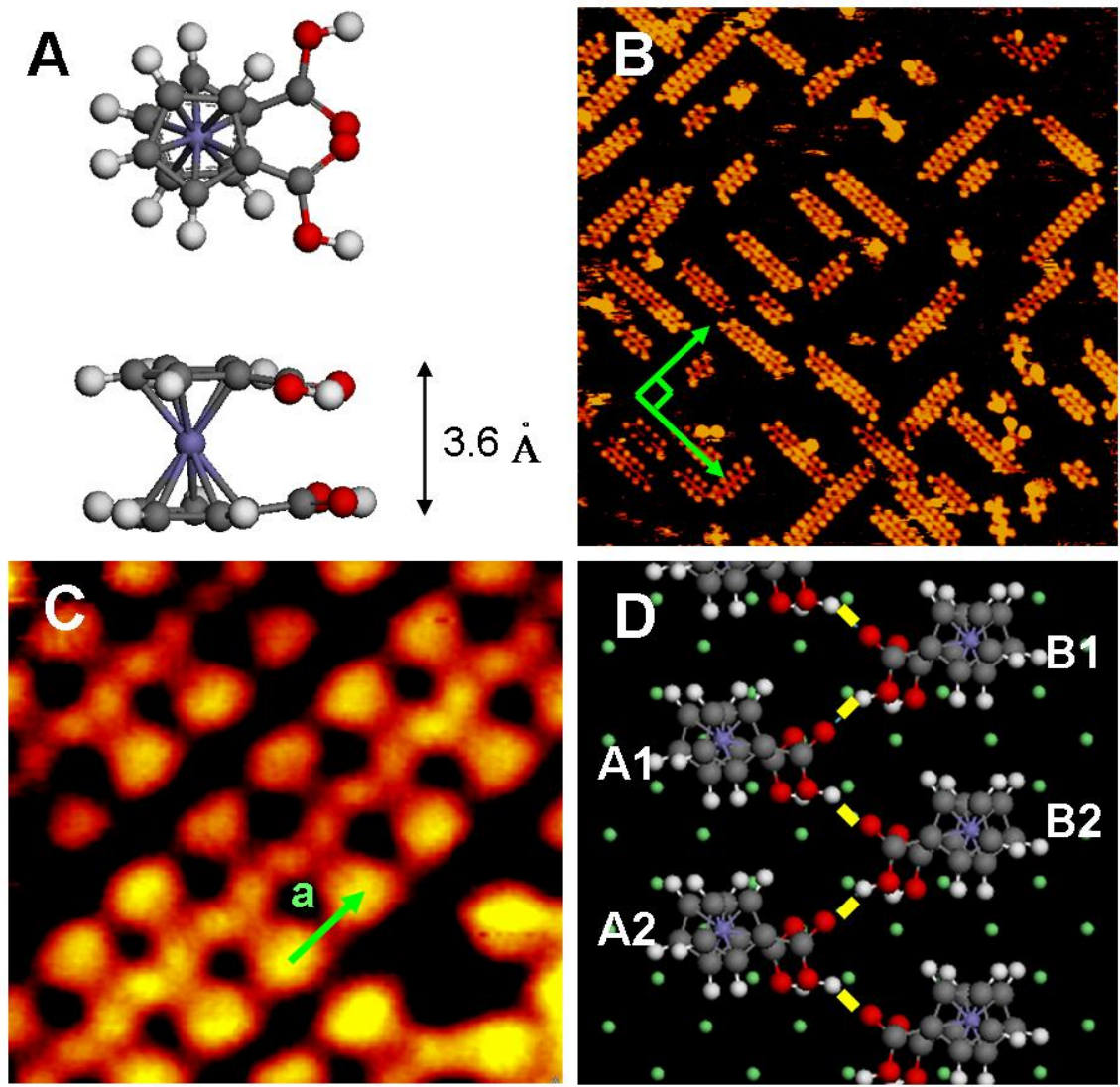

Figure 4

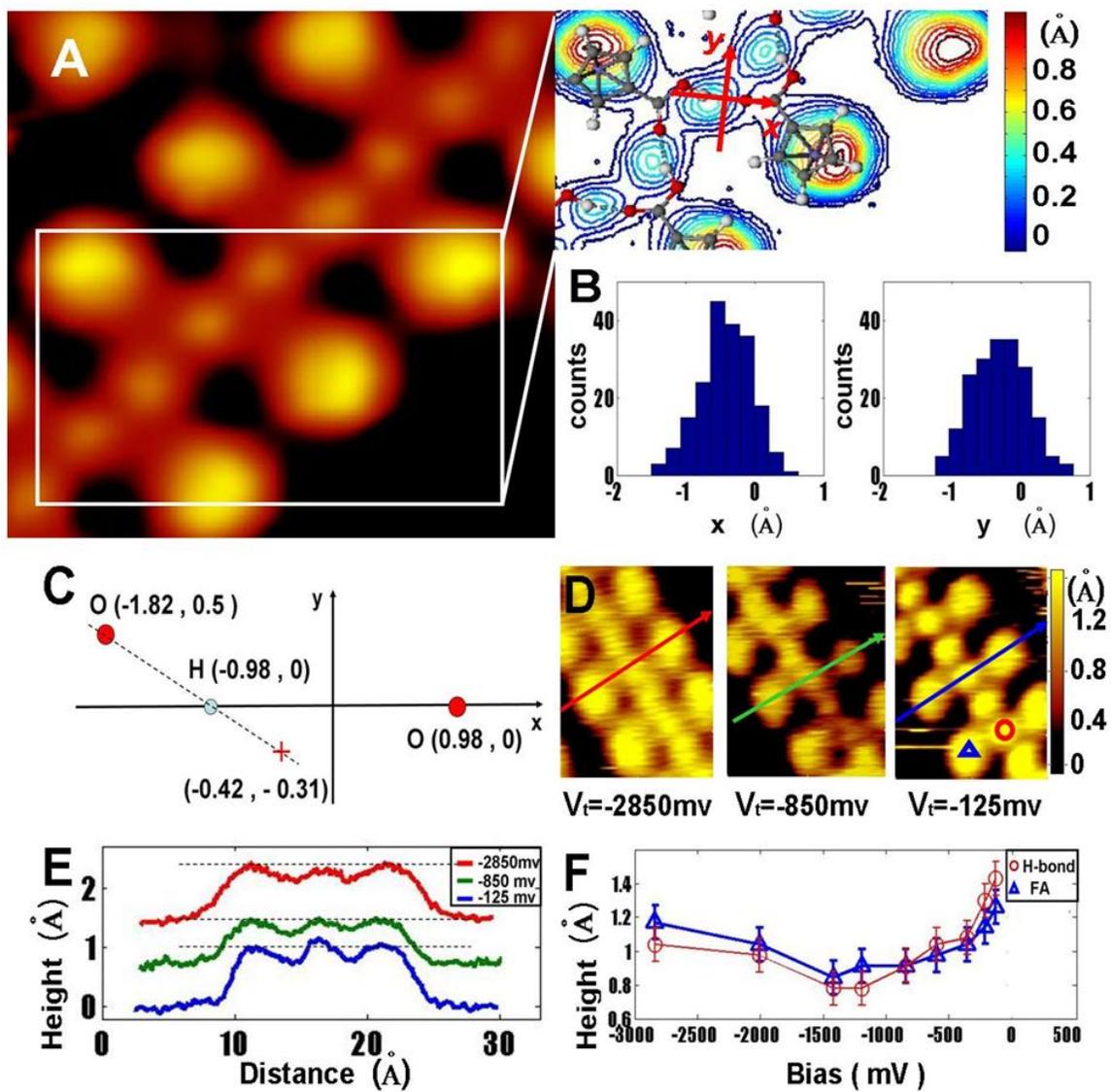

**Supplementary materials:**

**Methods**

**Experimental setup.** All the experiments were carried out in a multichamber with ultrahigh vacuum (UHV) system housing a SPECS variable temperature STM with base pressure of less than $2\times10^{-10}$ mbar. All STM images were acquired at room temperature with a chemically etched W tip. Positive voltage indicates that the samples were biased positively with respect to the tip.

**Sample preparation.** Gold single crystal (111) and Copper single crystal (100) (Mateck GmbH, Germany) were cleaned by repeated cycles of Ar+ bombardment (500 eV) and annealing (~800 K). After several hours of degassing, 1,3-Adamantanedicarboxylic Acid (Sigma-Aldrich, purity 98%) was evaporated from a Knudsen cell at 580 K onto a clean Au (111) surface held at room temperature. After degassing for several hours, 1,1'-Ferrocenedicarboxylic Acid (Tokyo Chemical Industry, purity 98% ) was also evaporated from a Knudsen cell held at 420 K onto a clean Cu(100) surface kept at room temperature.